\documentclass[a4paper,11pt]{article}

\usepackage[T1]{fontenc}
\usepackage[english]{babel}

\usepackage{palatino}
\usepackage[margin=25mm]{geometry}
\usepackage{amsmath,amssymb}

\newcommand*{\bs}{\boldsymbol}
\newcommand*{\mcal}{\mathcal}

\newcommand*{\diff}{\mathop{}\!\mathrm{d}}% Differential
\newcommand*{\pd}{\partial}% Partial differential
\newcommand*{\e}{\mathop{\mathrm{e}}\nolimits}% The number "e"
\newcommand*{\const}{\ensuremath{\mathrm{const}}}% const
\DeclareMathOperator{\grad}{\nabla}% Gradient
\DeclareMathOperator{\diverg}{\nabla\cdot}%{div\!}% Divergence
\newcommand*{\defin}{\stackrel{\mathrm{def}}{=}}
\newcommand*{\T}{\mathrm{T}}% Transpose
\newcommand*{\average}[1]{\langle#1\rangle}
\newcommand*{\Lapl}{\mathcal{L}}%{\tilde}% Laplace transform
\newcommand*{\R}{\mathbb{R}}
\newcommand*{\V}{\mathbb{V}}% Space of velocities
\newcommand*{\sphere}{\mathbb{S}}% "1D" sphere
\newcommand*{\outersol}{{\mathrm{o}}}
\newcommand*{\innersol}{{\mathrm{i}}}
\newcommand*{\aphase}{{\mathrm{a}}}
\newcommand*{\dphase}{{\mathrm{d}}}
\newcommand*{\eff}{{\mathrm{eff}}}

\numberwithin{equation}{section}

\usepackage[numbers, sort&compress]{natbib}
\newcommand*{\ie}{i.\,e.}

\newcommand*{\dd}{\ldots{}}

\usepackage{graphicx}
\usepackage{xcolor}
\usepackage{indentfirst}

\usepackage[unicode,colorlinks]{hyperref}

\definecolor{darkblue}{rgb}{0,0,0.8}

\hypersetup{%
  bookmarksopen=true,
  bookmarksopenlevel=1,
  bookmarksnumbered=true,
  linkcolor=darkblue,%darkred,
  citecolor=darkblue,%darkgreen,
  pdfpagelayout=OneColumn,
  pdfstartview={XYZ null null 1},%FitH,%
  pdfview={XYZ null null null}
}

\begin{document}

\title{%
  \bfseries
  A mean-field model of intermittent particle transport and its quasi-steady-state approximation
}

\author{%
  \bfseries
  \large
  Sergey~A.~Rukolaine\thanks{E-mail address: \texttt{rukol@ammp.ioffe.ru}}
}

\date{%
  % \itshape
  \small
  Ioffe Institute, 26 Polytekhnicheskaya, St.\,Petersburg 194021, Russia
}

\maketitle

\section{Introduction}
% \label{sec:Intro}

There are various models of intermittent particle transport (including intracellular transport)~\cite{Holcman:2007, LagacheEtAl:2009, BenichouEtAl:2011, ThielEtAl:2012, BressloffNewby:2013, Bressloff:2014}. We propose a mean-field model of intermittent particle transport, where a particle may be in one of two phases: the first is an active (ballistic) phase, when a particle runs with constant velocity in some direction, and the second is a passive (diffusive) phase, when the particle diffuses freely. The particle can instantly change the phase of motion. When the particle is in the active phase the rate of transition to the passive phase depends, in general, on time from the beginning of the run, so the distribution of the free path is not exponential. When the particle is in the passive phase the transition rate is constant, and diffusion is non-anomalous Brownian.
Reasoning is similar to that of Ref.~\cite{Rukolaine:2016}.

\section{Basic equations}

Particles (individuals) move in the $d$-dimensional space $\R^d$. They can be in one of two phases: the first is an active (or ballistic) phase, when a particle runs with constant velocity in some direction, and the second is a passive phase, when the particle diffuses freely. The particle can instantly change the phase of motion. When the particle is in the active phase the rate of transition to the passive phase depends, in general, on time from the beginning of the run. The densities of the particles in the active and passive phases are denoted by $\xi \equiv \xi(\bs{r}, t, \bs{v}, \tau)$ and $\vartheta \equiv \vartheta(\bs{r}, t)$, respectively, where $\bs{r}$ and $t$ are the space and time variables, respectively, $\bs{v} \in \V$ is the velocity vector, and $\tau$ is time from the beginning of the run.

The density $\xi$ obeys the equation%, see \cite{Alt:1980, Rukolaine:2016},
\begin{equation*}
  % \label{eq:PDEDensActivePhase}
  \pd_t^{} \xi
  + \pd_\tau^{} \xi
  + \bs{v} \cdot \grad\xi
  + \gamma \xi
  = 0,
\end{equation*}
where $\gamma \equiv \gamma(\bs{r}, \bs{v}, \tau)$ is the transition rate. This equation describes noninteracting particles running at the point $\bs{r}$ with the velocity $\bs{v}$ so that they become diffusing particles with the rate $\gamma$.

The density of particles at the beginning of the run is given by
\begin{equation*}
  % \label{eq:BoundCondDensActivePhase}
  \left. \xi \right|_{\tau=0}^{}
  =
  \alpha T(\bs{r}, \bs{v}) \vartheta,
\end{equation*}
where $\alpha \equiv \alpha(\bs{r})$ is the rate of transition from the passive to active phase, $T$ is the transition kernel such that
% \clr{$T \geq 0$ and $\int_\V^{} T(\bs{r}, \bs{v}) \diff\bs{v} = 1$.}
\begin{equation*}
  % \label{eq:TransitionKernelConditions}
  T \geq 0
  \quad\text{and}\quad
  \int_\V^{} T(\bs{r}, \bs{v}) \diff\bs{v}
  = 1.
\end{equation*}

The density $\vartheta$ obeys the diffusion equation
\begin{equation*}
  % \label{eq:PDEDensPassivePhase}
  \pd_t^{} \vartheta
  - D \Delta \vartheta
  =
  - \alpha \vartheta
  + \int_\V^{} 
  \int_0^\infty \gamma(\bs{r}, \bs{v}, \tau) \,\xi(\bs{r}, t, \bs{v}, \tau) \diff\tau \diff\bs{v},
\end{equation*}
where $D$ is the diffusion coefficient. The first term on the right-hand side expresses the loss of diffusing particles due to transition to the active phase, the second term is the source of diffusing particles due to transition from the active phase.

Initial conditions are
\begin{equation*}
  % \label{eq:InitCondDensActivePhase}
  \left. \xi \right|_{t=0}^{}
  =
  \chi(\bs{r}, \bs{v}) \delta(\tau),
\end{equation*}
where $\delta(\cdot)$ is the Dirac delta function, and
\begin{equation*}
  % \label{eq:InitCondDensPassivePhase}
  \left. \vartheta \right|_{t=0}^{}
  =
  \vartheta_0^{}(\bs{r}).
\end{equation*}
The initial condition for $\xi$ means that at the moment $t = 0$ all the active particles are at the beginning of the run.

\section{Transformed equations}

We define the density
\begin{equation*}
  \psi(\bs{r}, t, \bs{v})
  =
  \int_0^\infty  \xi(\bs{r}, t, \bs{v}, \tau) \diff \tau.
\end{equation*}

The densities $\psi$ and $\vartheta$ obeys the coupled equations
\begin{equation}
  \label{eq:PDEDensActivePhaseM}
  \pd_t^{} \psi
  + \bs{v} \cdot \grad\psi
  =
  - \mcal{M} \psi(\bs{r}, t, \bs{v})
  + \alpha T(\bs{r}, \bs{v}) \vartheta,
\end{equation}
% and
\begin{equation}
  \label{eq:PDEDensPassivePhaseM}
  \pd_t^{} \vartheta
  - D \Delta \vartheta
  =
  \int_\V^{} \mcal{M} \psi(\bs{r}, t, \bs{v}) \diff\bs{v}
  - \alpha \vartheta.
\end{equation}
The memory operator $\mcal{M}$ is given by
\begin{equation*}
  % \label{eq:MemoryOperator}
  \mcal{M} \psi(\bs{r}, t, \bs{v})
  =
  \int_0^t \phi(\bs{r}, \bs{v}, \tau) \,\psi(\bs{r} - \bs{v} \tau, t - \tau, \bs{v}) \diff \tau,
\end{equation*}
the memory kernel $\phi$ is defined through its Laplace transform by
\begin{equation*}
  % \label{eq:MemoryKernel}
  \Lapl \phi(\bs{r}, \bs{v}, s)
  \defin
  \frac{\Lapl p(\bs{r}, \bs{v}, s)}{\Lapl S(\bs{r}, \bs{v}, s)}
  % \Lapl p(\bs{r}, \bs{v}, s) / \Lapl S(\bs{r}, \bs{v}, s)
\end{equation*}
($\Lapl f(s) = \int_0^\infty \e^{-s \tau} f(\tau) \diff \tau$),
\begin{equation*}
  % \label{eq:SurvivalProb}
  S(\bs{r}, \bs{v}, \tau)
  =
  \exp \left\{ - \int_0^\tau \gamma(\bs{r} - \bs{v} (\tau -\tau'), \bs{v}, \tau') \diff \tau' \right\}
\end{equation*}
is the survival probability (the probability that the particle, started at the point $\bs{r} - \bs{v} \tau$ with the velocity $\bs{v}$, is still running),
\begin{equation*}
  % \label{eq:PDF}
  p(\bs{r}, \bs{v}, \tau)
  =
  \gamma(\bs{r}, \bs{v}, \tau) S(\bs{r}, \bs{v}, \tau),
\end{equation*}
$p(\bs{r}, \bs{v}, \tau) \diff \tau$ is the probability that the particle, started at the point $\bs{r} - \bs{v} \tau$, stops its run (becomes a diffusing particle) in the interval $(\bs{r}, \bs{r} + \bs{v} \diff \tau)$. [If $\gamma = \const$ then $p(\bs{v}, \cdot)$ is a probability density function, but, in general, this is not the case.]

The initial condition for the density $\psi$ (following from that for $\xi$) is
\begin{equation*}
  % \label{eq:InitCondDensActivePhaseM}
  \left. \psi \right|_{t=0}^{}
  =
  \chi(\bs{r}, \bs{v}).
\end{equation*}

Note that if $\gamma = \const$ then $\phi(\tau) = \gamma \delta(\tau)$ (\ie, particles in the active phase are memoryless, and the distribution $p(\cdot, \tau)$ is exponential) and, hence, $\mcal{M} \psi = \gamma \psi$. In this case Eqs.~\eqref{eq:PDEDensActivePhaseM} and \eqref{eq:PDEDensPassivePhaseM} are similar to equations in Ref.~\cite{BressloffNewby:2011}.

\section{Asymptotic solution}

\subsection{Assumptions}

\textbullet{} The transition rates $\alpha$ and $\gamma$ do not depend on $\bs{r}$. The transition rate $\gamma$ is also independent of the direction of the particle movement (\ie, $\gamma$ depends on the velocity $v$ rather than the direction $\bs\varOmega = \bs{v}/v$). Therefore the same is valid for the survival probability $S$, density $p$ and memory kernel $\phi$.

\textbullet{} The time of a run has the finite mean
\begin{equation*}
  \average{\tau_\aphase^{}}
  \equiv
  \average{\tau_\aphase^{}}(v)
  =
  \int_0^\infty \tau p(v, \tau) \diff\tau
\end{equation*}
and second moment
\begin{equation*}
  \average{\tau_\aphase^2}
  \equiv
  \average{\tau_\aphase^2}(v)
  =
  \int_0^\infty \tau^2 p(v, \tau) \diff\tau.
\end{equation*}

\textbullet{} The mean time of a run $\average{\tau_\aphase^{}}$ is small of order $\varepsilon$, where $\varepsilon$ is a small parameter, therefore the transition rate $\gamma$ is represented as
\begin{equation*}
  % \label{eq:ExtincCoeffEps}
  \gamma(v, \tau)
  =
  \frac{1}{\varepsilon} \,\bar\gamma \!\left( v, \frac{\tau}{\varepsilon} \right),
\end{equation*}
where $\bar\gamma$ does not depend on $\varepsilon$.
This implies the representations
\begin{equation*}
  S(v, \tau)
  =
  \bar{S} \!\left( v, \frac{\tau}{\varepsilon} \right),
  \quad
  p(v, \tau)
  =
  \frac{1}{\varepsilon} \,\bar{p} \!\left( v, \frac{\tau}{\varepsilon} \right)
  \quad\text{and}\quad
  \phi(v, \tau)
  =
  \frac{1}{\varepsilon^2} \,\bar\phi \!\left( v, \frac{\tau}{\varepsilon} \right),
\end{equation*}
where $\bar{S}$, $\bar{p}$ and $\bar\phi$ do not depend on $\varepsilon$.

\textbullet{} The mean time in the passive phase $\average{\tau_\dphase^{}}$ is small of order $\varepsilon$. Since $\average{\tau_\dphase^{}} = 1 / \alpha^{}$, this means that
\begin{equation*}
  \alpha
  =
  \frac{\bar\alpha}{\varepsilon},
\end{equation*}
where $\bar\alpha$ does not depend on $\varepsilon$.

\textbullet{} The diffusion coefficient $D$ is small of order $\varepsilon$:
\begin{equation*}
  D = \varepsilon \bar{D},
\end{equation*}
where $\bar{D}$ does not depend on $\varepsilon$.

\textbullet{} The transition kernel has the form
\begin{equation*}
  T(\bs{r}, \bs{v})
  =
  \bar{T}(\bs{r}, \bs{v})
  + \varepsilon \tilde{T}(\bs{r}, \bs{v}),
\end{equation*}
where $\bar{T}$ and $\tilde{T}$ do not depend on $\varepsilon$,
\begin{equation*}
  \bar{T} \geq 0
  \quad\text{and}\quad
  \int_{\V}^{} \bar{T}(\bs{r}, \bs{v}) \diff\bs{v}
  = 1.
\end{equation*}
We suppose also that
$\bar{T}(\bs{r}, -\bs{v}) = \bar{T}(\bs{r}, \bs{v})$, which is met if $\bar{T}$ is isotropic, \ie, $\bar{T} = |\V|^{-1}$.
The normalization of $T$ and $\bar{T}$ implies
\begin{equation*}
  \int_{\V}^{} \tilde{T}(\bs{r}, \bs{v}) \diff\bs{v}
  = 0.
\end{equation*}

\textbullet{} The ``space'' of velocities $\V$ is bounded and rotationally invariant.

\subsection{Asymptotic solution}

Eqs.~\eqref{eq:PDEDensActivePhaseM} and~\eqref{eq:PDEDensPassivePhaseM} are recast as
\begin{equation}
  \label{eq:DensPsiEps}
  \pd_t^{} \psi
  + \bs{v} \cdot \grad\psi
  =
  - \frac{1}{\varepsilon} \mcal{M}_\varepsilon^{} \psi(\bs{r}, t, \bs{v})
  + \frac{\bar\alpha}{\varepsilon} \left[ \bar{T}(\bs{r}, \bs{v}) + \varepsilon \tilde{T}(\bs{r}, \bs{v}) \right] \vartheta,
\end{equation}
\begin{equation}
  \label{eq:DensThetaEps}
  \pd_t^{} \vartheta
  - \varepsilon \bar{D} \Delta \vartheta
  =
  \frac{1}{\varepsilon} \int_\V^{} \mcal{M}_\varepsilon^{} \psi(\bs{r}, t, \bs{v}) \diff\bs{v}
  - \frac{\bar\alpha}{\varepsilon} \vartheta,
\end{equation}
where
\begin{equation*}
  \mcal{M}_\varepsilon^{} \psi(\bs{r}, t, \bs{v})
  =
  \int_0^t \frac{1}{\varepsilon} \bar\phi \!\left( v, \frac{\tau}{\varepsilon} \right) \psi(\bs{r} - \bs{v} \tau, t - \tau, \bs{v}) \diff \tau.
\end{equation*}

The densities are represented in the form
\begin{equation*}
  % \label{eq:DensPsiSum}
  \psi(\bs{r}, t, \bs{v})
  =
  \psi^\outersol(\bs{r}, t^\outersol, \bs{v})
  + \psi^\innersol(\bs{r}, t^\innersol, \bs{v}),
\end{equation*}
\begin{equation*}
  % \label{eq:DensThetaSum}
  \vartheta(\bs{r}, t)
  =
  \vartheta^\outersol(\bs{r}, t^\outersol)
  + \vartheta^\innersol(\bs{r}, t^\innersol),
\end{equation*}
where
\begin{equation*}
  t^\outersol = \varepsilon t
  \quad\text{and}\quad
  t^\innersol = t / \varepsilon
\end{equation*}
are slow time and fast time, respectively, $\psi^\outersol$, $\vartheta^\outersol$ and $\psi^\innersol$, $\vartheta^\innersol$ are outer and inner solutions, respectively. The inner solutions approximate $\psi$ and $\vartheta$ in the initial layer, \ie, for $0 < t \lesssim O(\varepsilon)$, while the outer solutions approximate them outside of the initial layer, \ie, for $t \gtrsim O(\varepsilon)$.
We derive here only the outer solutions. The inner solutions can be derived in the same way as in Ref.~\cite{Rukolaine:2016}.

The memory kernel has the asymptotic expansion%, see \cite{Rukolaine:2016},
\begin{equation*}
  % \label{eq:AsyExpansionMemoryKernel}
  \frac{1}{\varepsilon} \,\bar\phi \!\left( v, \frac{\tau}{\varepsilon} \right)
  \sim
  \bar\phi_0^{} \delta(\tau)
  + \bar\phi_1^{} \delta'(\tau) \varepsilon
  % + \bar\phi_2^{} \delta''(\tau) \varepsilon^2
  + \ldots{}
  \quad\text{as}\quad
  \varepsilon \to 0
\end{equation*}
(the expansion is considered in the weak sense)
with the coefficients
\begin{equation*}
  % \label{eq:CoeffMemoryKernelAsy}
  \bar\phi_0^{}
  \equiv
  \bar\phi_0^{}(v)
  =
  \frac{1}{\average{\bar\tau_\aphase^{}}},
  \quad%\text{and}\quad
  \bar\phi_1^{}
  \equiv
  \bar\phi_1^{}(v)
  =
  \frac{\average{\bar\tau_\aphase^2}}{2 \average{\bar\tau_\aphase^{}}^2} - 1,
\end{equation*}
% (an expression for $\bar\phi_2^{}$ is not needed here),
where
\begin{equation*}
  \average{\bar\tau_\aphase^{}}
  \equiv
  \average{\bar\tau_\aphase^{}}(v)
  =
  \int_0^\infty \tau \bar{p}(v, \tau) \diff\tau
  \quad\text{and}\quad
  \average{\bar\tau_\aphase^2}
  \equiv
  \average{\bar\tau_\aphase^2}(v)
  =
  \int_0^\infty \tau^2 \bar{p}(v, \tau) \diff\tau.
\end{equation*}

% It is convenient to recast Eqs.~\eqref{eq:DensPsiOuterEps} and~\eqref{eq:DensThetaOuterEps} for later use in the matrix-operator form, taking into account the expansion~\eqref{eq:MemoryKernelEps}, as
% \begin{equation}
%   \label{eq:DensOuterEps}
%   \left(
%     \varepsilon^2 \mcal{D}
%     + \varepsilon \mcal{V}
%   \right)
%   \begin{pmatrix}
%     \psi^\outersol\\[1ex]
%     \vartheta^\outersol
%   \end{pmatrix}
%   =
%   \left[
%     - \left( \Id - \mcal{T} \right) \mcal{B}_0^{}
%     + \varepsilon \mcal{A}_1^{}
%     + \varepsilon^2 \mcal{A}_2^{}
%     + \ldots
%   \right]
%   \begin{pmatrix}
%     \psi^\outersol\\[1ex]
%     \vartheta^\outersol
%   \end{pmatrix},
% \end{equation}

We assume that the outer solutions have the asymptotic expansions
\begin{equation*}
  \psi^\outersol
  \sim
  \psi_0^\outersol
  +
  \psi_1^\outersol \varepsilon
  +
  \psi_2^\outersol \varepsilon^2
  + \dd
  \quad\text{as}\quad
  \varepsilon \to 0,
\end{equation*}
\begin{equation*}
  \vartheta^\outersol
  \sim
  \vartheta_0^\outersol
  +
  \vartheta_1^\outersol \varepsilon
  +
  \vartheta_2^\outersol \varepsilon^2
  + \dd
  \quad\text{as}\quad
  \varepsilon \to 0.
\end{equation*}

% Substituting the asymptotic expansions~\eqref{eq:OuterAsyExpansion} into Eq.~\eqref{eq:DensOuterEps} and equating coefficients of like powers of $\varepsilon$ yields the following equations:
% \begin{subequations}
%   \label{eq:EqEps}
%   \begin{align}
%     \label{eq:EqEpsZero}
%     & \varepsilon^0:
%     \quad
%     \left( \Id - \mcal{T} \right) \mcal{B}_0^{}
%     \begin{pmatrix}
%       \psi_0^\outersol\\[1ex]
%       \vartheta_0^\outersol
%     \end{pmatrix}
%     =
%     \begin{pmatrix}
%       0\\[1ex]
%       0
%     \end{pmatrix},
%     \\
%     \label{eq:EqEpsOne}
%     & \varepsilon^1:
%     \quad
%     \left( \Id - \mcal{T} \right) \mcal{B}_0^{}
%     \begin{pmatrix}
%       \psi_1^\outersol\\[1ex]
%       \vartheta_1^\outersol
%     \end{pmatrix}
%     =
%     \left( - \mcal{V} + \mcal{A}_1^{} \right)
%     \begin{pmatrix}
%       \psi_0^\outersol\\[1ex]
%       \vartheta_0^\outersol
%     \end{pmatrix},
%     \\
%     \label{eq:EqEpsTwo}
%     & \varepsilon^2:
%     \quad
%     \left( \Id - \mcal{T} \right) \mcal{B}_0^{}
%     \begin{pmatrix}
%       \psi_2^\outersol\\[1ex]
%       \vartheta_2^\outersol
%     \end{pmatrix}
%     =
%     \left( - \mcal{D} + \mcal{A}_2^{} \right)
%     \begin{pmatrix}
%       \psi_0^\outersol\\[1ex]
%       \vartheta_0^\outersol
%     \end{pmatrix}
%     + \left( - \mcal{V} + \mcal{A}_1^{} \right)
%     \begin{pmatrix}
%       \psi_1^\outersol\\[1ex]
%       \vartheta_1^\outersol
%     \end{pmatrix},
%     \\
%     \notag
%     & \phantom{\varepsilon^2: \quad}
%     \cdots
%   \end{align}
% \end{subequations}

Substituting the asymptotic expansions into Eqs.~\eqref{eq:DensPsiEps}, \eqref{eq:DensThetaEps}, and equating coefficients of like powers of $\varepsilon$ yields the system of equations of the second kind for $\psi_i^\outersol$ and $\vartheta_i^\outersol$.
The use of the Fredholm alternative and the solvability condition yields the representation of the zeroth-order densities (the quasi-steady state approximation)
\begin{equation*}
  \begin{pmatrix}
    \psi_0^\outersol\\[1ex]
    \vartheta_0^\outersol
  \end{pmatrix}
  =
  \begin{pmatrix}
    \average{\bar\tau_\aphase^{}} \bar{T}(\bs{r}, \bs{v})\\[1ex]
    \average{\bar\tau_\dphase^{}}
  \end{pmatrix}
  \rho(\bs{r}, t^\outersol),
\end{equation*}
where $\average{\bar\tau_\dphase^{}} = 1 / \bar\alpha^{}$.

% The solvability condition implies
% \begin{equation*}
%   \pd_{t^\outersol}^{} \rho
%   - \diverg \left( \bar{D}_\eff^{} \grad \rho \right)
%   + \bar{\bs{u}} \cdot \grad \rho
%   = 0,
% \end{equation*}
% where
% \begin{equation*}
%   \bar{C}
%   =
%   \int_\V^{} \average{\bar\tau_\aphase^{}} \bar{T}(\bs{r}, \bs{v}) \diff\bs{v}
%   + \average{\bar\tau_\dphase^{}},
% \end{equation*}
% \begin{equation*}
%   \bar{D}_\eff^{}
%   =
%   \frac{1}{\bar{C}} \left[
%     \bar{D} \average{\bar\tau_\dphase^{}}
%     + \int_\V^{} \frac{\average{\bar\tau_\aphase^{2}}}{2} \bar{T}(\bs{r}, \bs{v}) \bs{v} \bs{v}^\T \diff\bs{v}
%   \right]
% \end{equation*}
% and
% \begin{equation*}
%   \bar{\bs{u}}
%   =
%   \frac{1}{\bar{C}} \int_\V^{} \average{\bar\tau_\aphase^{}} \tilde{T}(\bs{r}, \bs{v}) \bs{v} \diff\bs{v}
% \end{equation*}

The density $\rho \equiv \rho(\bs{r}, t)$ satisfies the drift-diffusion equation
\begin{equation*}
  \pd_t^{} \rho
  - \diverg \left( D_\eff^{} \grad \rho \right)
  + \bs{u} \cdot \grad \rho
  = 0,
\end{equation*}
where
\begin{equation*}
  C
  =
  \int_\V^{} \average{\tau_\aphase^{}} \bar{T}(\bs{r}, \bs{v}) \diff\bs{v}
  + \average{\tau_\dphase^{}},
\end{equation*}
the effective diffusion coefficient is
\begin{equation*}
  D_\eff^{}
  =
  \frac{1}{C} \left[
    D \average{\tau_\dphase^{}}
    + \int_\V^{} \frac{\average{\tau_\aphase^{2}}}{2} \bar{T}(\bs{r}, \bs{v}) \bs{v} \bs{v}^\T \diff\bs{v}
  \right]
\end{equation*}
the drift velocity is
\begin{equation*}
  \bs{u}
  =
  \frac{1}{C} \int_\V^{} \average{\tau_\aphase^{}} \left[ T(\bs{r}, \bs{v}) - \bar{T}(\bs{r}, \bs{v}) \right] \bs{v} \diff\bs{v}.
\end{equation*}

The initial condition is
\begin{equation*}
  \rho|_{t=0}
  =
  \int_\V \bar{T}(\bs{r}, \bs{v}) \bar\phi_0^{}(v) \chi(\bs{r}, \bs{v}) \diff\bs{v}
  + \bar\alpha(\bs{r}) \vartheta_0^{}(\bs{r}).
\end{equation*}

\subsection{A special case: the one-speed model}

Let $v \equiv |\bs{v}| = \const$ (\ie, $\V = v \,\sphere^{d-1}$) and the nonperturbed term $\bar{T}$ of the transition kernel is isotropic, \ie,
$\bar{T}(\bs{r}, \bs\varOmega) = |\V|^{-1}$,
% \begin{equation*}
%   \bar{T}(\bs{r}, \bs\varOmega)
%   =
%   \frac{1}{|\V|}
%   %   \equiv
%   %   \frac{1}{v^{d-1} |\sphere^{d-1}|}
% \end{equation*}
then $C = \average{\tau_\aphase^{}} + \average{\tau_\dphase^{}}$, and
% \begin{equation*}
%   C
%   =
%   \average{\tau_\aphase^{}}
%   + \average{\tau_\dphase^{}},
% \end{equation*}
\begin{equation*}
  D_\eff^{}
  =
  \frac{1}{\average{\tau_\dphase^{}} + \average{\tau_\aphase^{}}}
  \left[
    D \average{\tau_\dphase^{}} + \frac{\average{\tau_\aphase^2} v^2}{2 d}
  \right],
\end{equation*}
\begin{equation*}
  \bs{u}
  =
  \frac{\average{\tau_\aphase^{}} v}{\average{\tau_\dphase^{}} + \average{\tau_\aphase^{}}} \int_{\sphere^{d-1}}^{} \left[ T(\bs{r}, \bs\varOmega) - \bar{T}(\bs{r}, \bs\varOmega) \right] \bs\varOmega \diff\bs\varOmega.
\end{equation*}

Limiting cases:
% \begin{equation*}
%   D_\eff^{}
%   \to
%   \begin{cases}
%     D
%     & \text{as}\quad \average{\tau_\aphase^{}} \to 0\\[1ex]
%     \dfrac{\average{\tau_\aphase^2} v^2}{2 d \average{\tau_\aphase^{}}}
%     & \text{as}\quad \average{\tau_\dphase^{}} \to 0
%   \end{cases}
% \end{equation*}
% \begin{equation*}
%   \bs{u}
%   \to
%   \begin{cases}
%     0
%     & \text{as}\quad \average{\tau_\aphase^{}} \to 0\\[1ex]
%     v \displaystyle\int_{\sphere^{d-1}}^{} \left[ T(\bs{r}, \bs\varOmega) - \bar{T}(\bs{r}, \bs\varOmega) \right] \bs\varOmega \diff\bs\varOmega
%     & \text{as}\quad \average{\tau_\dphase^{}} \to 0
%   \end{cases}
% \end{equation*}
If the active phase is negligeably short compared to the passive phase ($\average{\tau_\aphase^{}} \ll \average{\tau_\dphase^{}}$), then
\begin{equation*}
  D_\eff^{}
  \to
  D
  \quad\text{and}\quad
  \bs{u}
  \to 0.
\end{equation*}
If the passive phase is negligeably short compared to the active phase ($\average{\tau_\dphase^{}} \ll \average{\tau_\aphase^{}}$), then
\begin{equation*}
  D_\eff^{}
  \to
  \dfrac{\average{\tau_\aphase^2} v^2}{2 d \average{\tau_\aphase^{}}}
  \quad\text{and}\quad
  \bs{u}
  \to
  v \displaystyle\int_{\sphere^{d-1}}^{} \left[ T(\bs{r}, \bs\varOmega) - \bar{T}(\bs{r}, \bs\varOmega) \right] \bs\varOmega \diff\bs\varOmega
\end{equation*}

\subsection{The importance of being corrected}

A correction to the diffusion coefficient $D_\eff^{}$ due to non-exponential distribution of free paths may be significant.
Consider the one-speed model. Suppose for simplicity that $\average{\tau_\dphase^{}} = 0$ (the passive phase is negligeably short) and $T(\bs{r}, \bs\varOmega) = \bar{T}(\bs{r}, \bs\varOmega) = |\V|^{-1}$ (reorientation is isotropic). Then $\bs{u} = 0$ and $D_\eff^{} = \average{\tau_\aphase^2} v^2 / (2 d \average{\tau_\aphase^{}})$.

\begin{figure}[htb]
  \includegraphics[width=\linewidth]{%
    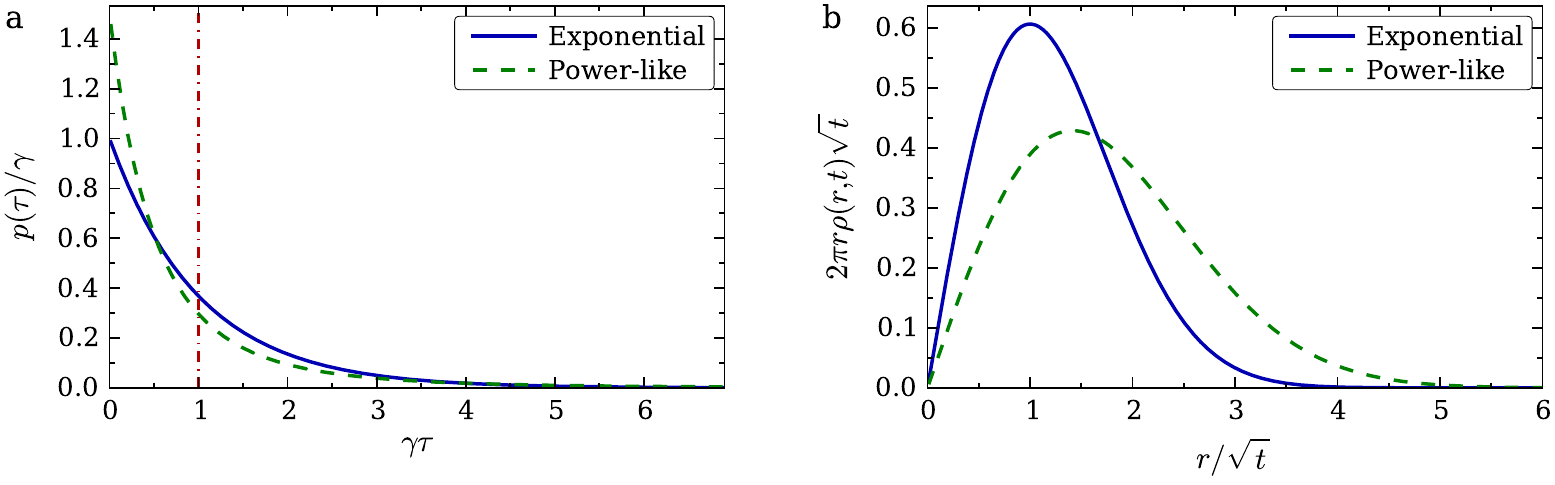}
  \caption{(a) The exponential and power-like PDFs with the same mean (shown by the vertical dash-dot line). (b) Corresponding solutions to the initial value problem~\eqref{eq:DiffusionEq}, \eqref{eq:InitCond}, $d=2$.}
  \label{fig:Run_time_distributions_and_diffusion_d2}
\end{figure}

\begin{figure}[htb]
  \includegraphics[width=\linewidth]{%
    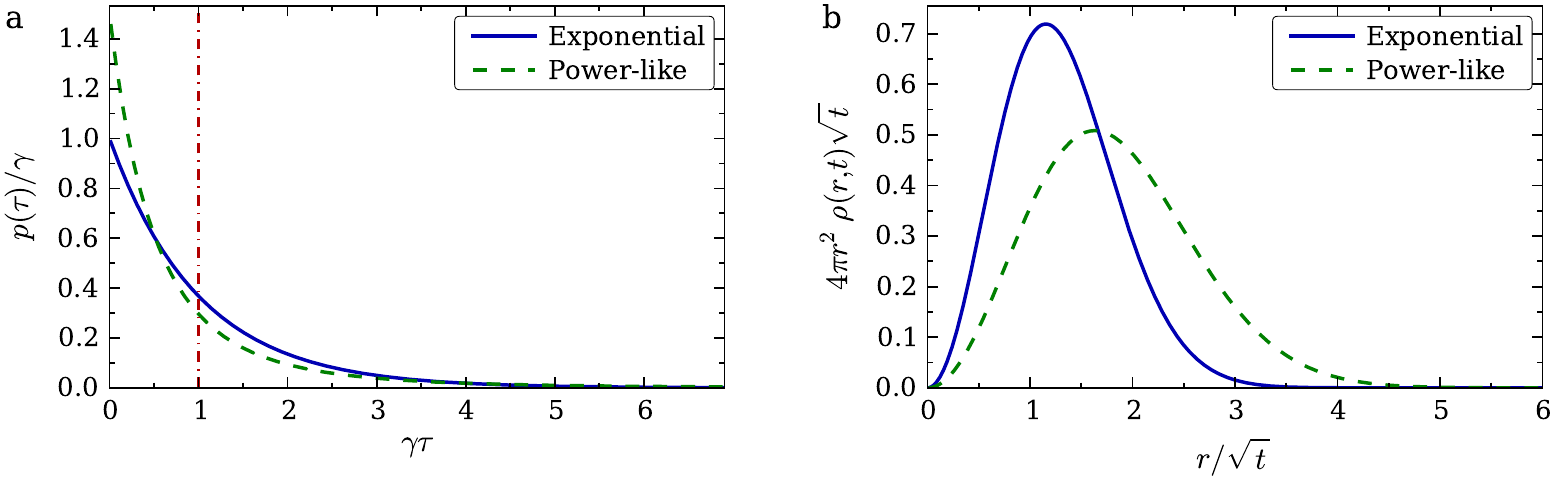}
  \caption{(a) The exponential and power-like PDFs with the same mean (shown by the vertical dash-dot line). (b) Corresponding solutions to the initial value problem~\eqref{eq:DiffusionEq}, \eqref{eq:InitCond}, $d=3$.}
  \label{fig:Run_time_distributions_and_diffusion_d3}
\end{figure}

Fig.~\ref{fig:Run_time_distributions_and_diffusion_d2}(a) (Fig.~\ref{fig:Run_time_distributions_and_diffusion_d3}(a) is the same) shows two probability density functions (PDFs): the first is the exponential PDF
\begin{equation*}
  p(\tau)
  =
  \gamma \e^{-\gamma \tau}
\end{equation*}
and the second is the power-like PDF
\begin{equation*}
  p(\tau)
  =
  \frac{\gamma \alpha}{\alpha - 1} \left[ 1 + \frac{\gamma \tau}{\alpha - 1} \right]^{- (\alpha + 1)}
  % \propto
  % \tau^{-(\alpha + 1)}\\
  % \quad\text{as}\quad
  % \tau \to \infty,
\end{equation*}
[$p(\tau) \propto \tau^{-(\alpha + 1)}$ as $\tau \to \infty$], $\alpha = 3$. Both distributions have the same mean $\average{\tau_\aphase^{}} = 1 / \gamma$.

Figs.~\ref{fig:Run_time_distributions_and_diffusion_d2}(b) and \ref{fig:Run_time_distributions_and_diffusion_d3}(b) show solutions of the diffusion equation in the two- and three-dimensional spaces
\begin{equation}
  \label{eq:DiffusionEq}
  \pd_t^{} \rho
  - D_\eff^{} \Delta \rho
  = 0,
  \quad
  \bs{r} \in \R^d,
\end{equation}
with the initial condition
\begin{equation}
  \label{eq:InitCond}
  \left. \rho \right|_{t=0}^{}
  =
  \delta(\bs{r}),
\end{equation}
for the two PDFs ($v=1$).
(The solutions remain unchanged in the given scales.)
The diffusion coefficient $D_\eff^{}$ corresponding to the power-like PDF is twice as large as the diffusion coefficient corresponding to the exponential PDF.


\section{Concluding remarks}
\label{sec:Concl}

\begin{list}{\textbullet{}}{\topsep=1ex \partopsep=0pt \parskip=0pt \parsep=0pt \itemsep=0pt \labelwidth=1ex \labelsep=0.5ex \leftmargin=1.5ex}

\item%[\textbullet{}] 
  The use of the exponential distribution of free paths, when it is actually non-exponential, may lead to significant errors in results of modelling.

  % \item%[\textbullet{}] 
  %   Non-exponential distribution of free path lengths should be taken into account to avoid significant errors in results of modelling.

\item%[\textbullet{}] 
  The model admits straightforward extension to the case when particles are absorbed (degraded) and there are sources of particles.

\item%[\textbullet{}] 
  Extension of the model to non-Cartesian geometries is possible. For example, it can be extened to intracellular vesicular transport in spherical geometry.

\end{list}

% \bibliographystyle{unsrt}
% \bibliography{TMCSLS_2017_poster_arXiv}

\end{document}